\documentclass[twocolumn,showpacs, showkeys,preprintnumbers,amsmath,amssymb]{revtex4}

\usepackage{graphicx}
\usepackage{dcolumn}
\usepackage{bm}

\begin{document}

\preprint{draft}

\title{Upper and lower limits on the Crab pulsar's astrophysical parameters set from gravitational wave observations by LIGO: 
\\ braking index and energy considerations.} 

\author{Giovanni Santostasi}

\email{gsantostasi@mcneese.edu}
\affiliation{ Physics Department, McNeese State University 
}%

\date{\today}
             
\begin{abstract}
\begin{center}
\bf{ \normalsize{Abstract} }
\end{center}
The Laser Interferometer Gravitational Observatory (LIGO) has recently reached the end of its fifth science run (S5), having collected more than a year worth of data. Analysis of the data is still ongoing but a positive detection of gravitational waves, while possible, is not realistically expected for most likely sources. This is particularly true for what concerns gravitational waves from known pulsars. In fact, even under the most optimistic (and not very realistic) assumption that all the pulsar's observed spin-down is due to gravitational waves, the gravitational wave strain at earth from all the known isolated pulsars (with the only notable exception of the Crab pulsar) would not be strong enough to be detectable by existing detectors. By August 2006, LIGO had produced enough data for a coherent integration capable to extract signal from noise that was weaker than the one expected from the Crab pulsar's spin-down limit. No signal was detected, but beating the spin-down limit is a considerable achievement for the LIGO Scientific Collaboration (LSC). It is customary to translate the upper limit on strain from a pulsar into a more astrophysically significant upper limit on ellipticity. Once the spin-down limit has been beaten, it is possible to release the constraint that all the spin-down is due to gravitational wave emission. A more complete model with diverse braking mechanisms can be used to set limits on several astrophysical parameters of the pulsar. This paper shows possible values of such parameters for the Crab pulsar given the current limit on gravitational waves from this neutron star.

\end{abstract}

\maketitle

\section{Introduction}
The LIGO observatory has conducted five science runs \cite{c1}, \cite{c8}, \cite{c9}. A coherent integration to extract signal from noise from possible continuous sources, such as lumpy rotating neutron stars, (using a technique explained in \cite{c10}) was performed.\\ To aid the search it is useful to implement filters matched to the signal derived from known neutrons' stars parameters such as the celestial coordinates, the rotation frequency and the frequency derivatives of the searched for source. Known pulsars are very convenient possible sources because such parameters are well known from radio and X-rays observations. No signal was detected from any known pulsars during LIGO's first four data runs. The coherent search was limited mainly to sinusoidal signals at frequencies that were twice the rotation frequency of the pulsar, typical of a non-axisymmetric rotating neutron star.\\ Analysis of the data is still ongoing for S5 but a positive detection of gravitational waves, while possible, is not realistically expected from any known pulsar. In fact, it is well known that even under the very optimistic and unrealistic assumption that all the spin-down is attributable to the emission of gravitational waves, the gravitational strain of such radiation reaching the earth would not be strong enough to be detectable by existing detectors, even after one year of integration time. It is common to illustrate this particular fact in a way similar to Fig.1. The maximum possible strain from the known pulsars is shown as a function of emission frequency and compared with the noise level for the an existing LIGO detector and planned upgrade. The sensitivity curve is based on the reduced noise level achieved after a full year of coherent integration (an improvement that is proportional to the square root of the integration time). Almost all the pulsars have maximum possible strains that are below the sensitivity curve and therefore are undetectable by the present LIGO. The maximum possible strain is an upper limit calculated from simple energy conservation arguments and the value of the observed spin-down of the pulsar. The only pulsars with maximum possible strains above the curve are the Crab pulsar (PSR B0531+21) and the pulsar J1022+1001 that is a member of a binary system (the analysis of the gravitational waves from this kind of source is more complicated and it will be not considered in this letter).\\ The spin-down limit for a known pulsar can be calculated as in the following. 
Under the optimistic assumption that all the spin-down is due solely to the emission of gravitational radiation, the pulsar loses energy with a rate given by:
\begin{equation}
\dot{E}=\frac{32}{5}\frac{G}{c^{5}}I^{2}\epsilon^{2}\Omega^{6},
\end{equation}
where $\Omega=2\pi f_{r}$, $f_{r}$ is the rotation frequency, $G$ is the gravitational constant, $c$ is the velocity of light, $I$ is the moment of inertia and $\epsilon$ is the pulsar ellipticity.
The radiated energy depletes the rotational kinetic energy reservoir by an amount $\dot{E}=\Omega\dot{\Omega}I$. The radiation emitted will be in the form of gravitational waves (at twice the rotation frequency for a simple non-axisymmetric rotator) with strain amplitude $h$ that can be determined from the time derivative of the energy. In fact, we have that:
\begin{equation}
h\approx\left(\frac{G}{\pi^2 c^3}\right)^{1/2}\frac{\dot{E}^{1/2}}{r f},
\end{equation}
where $f=2f_{r}$ is the gravitational wave frequency and $r$ is the distance of the source.\\ For the majority of known pulsars the angular frequency $\Omega$, the angular frequency time derivative $\dot{\Omega}$ and the distance $r$ are known from radio or X-ray observations. Using these values, it is common in the literature to illustrate the spin-down limits as shown in Figure 1.\\ The strain spin-down limit for the Crab Pulsar corresponds to a value of $h\leq1.4\times10^{-24}$.
The spin-down limit can be converted into a limit on the ellipticity of the pulsar according to the following equation:
\begin{equation}
\epsilon=9.5\times10^{-6}\left(\frac{r}{kpc}\right)\left(\frac{f}{kHz}\right)^{-2}\left(\frac{h}{10^{-23}}\right)\left(\frac{10^{45}gcm^{2}}{I_{zz}}\right)
\end{equation}
For the Crab pulsar the spin-down limit on ellipticity is $\epsilon=7.5\times10^{-4}$.\\
The Crab pulsar's emission frequency is very close to the power grid typical cycle. Resonances and possible frequency up-conversions make the noise floor near the Crab's emission frequency higher than what it would be without the additional noise due to the power grid interference. But enough integration time allowed the LIGO detector to reach and beat the spin-down limit. By August 2006 a preliminary upper limit value for the strain $h=5.0\times10^{-25}$ that corresponds to an upper limit on the ellipticity $\epsilon=2.5\times10^{-4}$ has been claimed by the LSC. This result is based on a search that used data up to August 25th 2006 when a major glitch occurred in the Crab pulsar's timing (creating a possible abrupt change of phase) and interrupted momentarily the analysis \cite{c11}. Recently, the results of more accurate and complete searches were announced \cite{c17}. Three searches, with different probability methods and physical assumptions, were performed giving a range of upper limits on strain from $1.7\times10^{-24}$ to $2.7\times10^{-25}$, which translates in a range of ellipticities values in the range from $9\times10^{-4}$ to $1.4\times10^{-4}$. In this paper we mainly use the single template limit on ellipticity $\epsilon=1.8\times10^{-4}$. This is a somehow conservative value because it doesn't depend on assumptions made on the orientation of the neutron star's spin axis.
 \begin{figure}
	\centering	\includegraphics[width=0.45\textwidth]{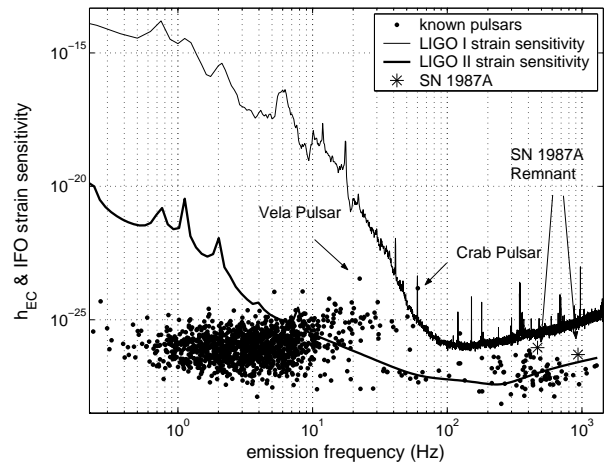}
	\caption{The graph shows the gravitational strain $h_{EC}$ (EC stands for Energy Conservation or spin-down limit) for the known pulsars from the complete ATNF catalog (about 1500 stars). The two star marks indicate the $h$ upper limit for the putative supernova compact remnant SN 1987A (being a precessing neutron star it emits a the rotation frequency and twice the rotation frequency)\cite{c5}. The Crab pulsar is above the noise curve even for the initial LIGO case. Unfortunately, it is very close to the 60 Hz line (caused by resonance with the power grid typical cycle.) The quantity $h_{EC}$ was derived assuming that all observed spin-down is due to energy loss caused by emission of gravitational radiation (and no other braking mechanisms.) The strain for the pulsars is compared with the noise sensitivity curve (in units of $h\sqrt{Hz}$) for the current LIGO detector (actual data for the Hanford detector taken during the science run S2 conducted in 2004 and extrapolated to 1 year of integration time). The expected noise curve for Advanced LIGO was computed numerically using the free-share code BENCH \cite{c13}.}	
\label{fig:specTOT}
\end{figure}

\section{Ratio of Energy Loss due to Gravitational versus electromagnetic Radiation}
\subsection{A simple model with mixed breaking mechanisms}
Once the spin-down limit for a particular pulsar has been reached, we can conclude, that in the absence of a positive detection of gravitational waves, the star is spinning down through a combination of braking mechanisms.
For example \cite{c3}, we can adopt a model that implies gravitational and electromagnetic braking. In this case the energy loss will be given by:
\begin{equation}
\dot{E}=I\Omega\dot{\Omega}=I\Omega\left(\dot{\Omega}_{GW}+\dot{\Omega}_{em}\right)	.
\end{equation}
The contribution to the energy loss due to gravitational waves is given by:
\begin{equation}
\dot{E}_{GW}=I\Omega\dot{\Omega}_{GW}=\frac{32}{5}\frac{G}{c^{5}}I^{2}\epsilon^{2}\Omega^{6},
\end{equation}
while the contribution to the energy loss due to dipole radiation and other electromagnetic processes is given by:
\begin{equation}
\dot{E}_{em}=I\Omega\dot{\Omega}_{em}.
\end{equation}
In general the spin-down can be expressed as a power law in terms of rotation frequency $\Omega$ as:
\begin{equation}
\dot{\Omega}=K\Omega^{n}	
\end{equation}
where $K$ is the torque function (the quantity $I\dot{\Omega}$ is the torque acting on the star) and $n$ is the braking index, that is telling of the mechanisms at work in spinning down the pulsar. For example, in the case of a pure electromagnetic dipole $n=3$ and $n=5$ for pure gravitational radiation (this last result can be easily derived using Eq. (5)). In the case of a mixed contribution of braking mechanisms, n will be different from the above values. In this case we will have:
\begin{equation}
\dot{\Omega}=K\Omega^{n}=	K_{em}\Omega^{n_{em}}+K_{GW}\Omega^{n_{GW}},
\end{equation}
where $K_{em}$ and $n_{em}$ are the electromagnetic torque function and braking index and $K_{GW}$ and $n_{GW}$ are the equivalent gravitational quantities.
In principle, the power law $n$ can be calculated from observational parameters of the pulsar.  In fact, using Eq. (4), we can derive the expression:
\begin{equation}
n=\frac{\Omega\ddot{\,\Omega}}{\,\dot{\Omega}^{2}}.
\end{equation}
Just for few pulsars (about four) we can measure the value of $\ddot{\,\Omega}$ reliably, therefore a calculation of $n$ from observable quantities is usually unpractical. \\
The Crab pulsar is one of such pulsars for which the value of $n$ is measured and it is equal to $2.51$. This allows us to set some interesting upper limits on the energetics of the pulsar and on the index of the braking mechanisms at work.\\
Following \cite{c3}, we can define the ratio $Y(\Omega )$ between the energy loss due to gravitational waves and the energy loss due to electromagnetic processes as:
\begin{equation}
Y(\Omega)=\frac{\dot{\Omega}_{GW}}{\dot{\Omega}_{em}}.
\end{equation} 
The braking index $n$ can be expressed in terms of the quantity Y as:
\begin{equation}
n=\frac{n_{em}+5Y}{1+Y},
\end{equation}and then solving for Y as a function of n we have:
\begin{equation}
Y(n)=\frac{n-n_{em}}{5-n}.
\end{equation}
Using Eq. (5), (10) and (12) we can calculate an upper limit on $Y(n)$ from the upper limit on ellipticity using the following expression:
\begin{equation}
\epsilon=7.55\times10^6\sqrt{\frac{|\dot{\Omega}|}{\Omega^{5}}\frac{Y(n)}{1+Y(n)}}
\end{equation}
then solving for $Y(n)$:
\begin{equation}
Y(n)=\frac{A}{1-A}
\end{equation}
where the expression $A$ is equal to:
\begin{equation}
A=\frac{\epsilon}{5.70\times10^{13}}\frac{\Omega^{5}}{|\dot{\Omega}|}.
\end{equation}
Finally, we can calculate a lower limit on the electromagnetic braking index $n_{em}$ using Eq. (12):
\begin{equation}
n_{em}=n\left(1+Y_{n}\right){5-n}.
\end{equation}
We can also write an expression for the ratio of the energy contribution of the electromagnetic and gravitational braking to the total energy loss $\dot{E}_{tot}$:
\begin{equation}
\frac{\dot{E}_{em}}{\dot{E}_{tot}}=\frac{1}{1+Y_{n}},
\end{equation}
and
\begin{equation}
\frac{\dot{E}_{GW}}{\dot{E}_{tot}}=\frac{Y_{n}}{1+Y_{n}},
\end{equation}
Using the Crab pulsar observational parameters $n=2.51$, $\Omega=186.96$ $ s^{-1}$, $\dot{\Omega}=-2.33\times10^{-9} s^{-2}$ and latest ellipticity limit $\epsilon=1.8\times10^{-4}$ from the LSC, we obtain current values for the upper and lower limits for the ratio $Y_{n}$, the electromagnetic braking index $n_{em}$ and an upper limit on the contribution of gravitational radiation to the total energy loss as illustrated in Table 1.

\begin{table}
\centering
\begin{tabular}[b] { |c|c|c| } 

\hline 
\normalsize{$Y_{n}$}& 
\normalsize{ 0.058} & 
\normalsize{upper limit}  	\\
\hline 
\normalsize{$n_{em}$} & 
\normalsize{2.36} & 
\normalsize{lower limit}  	\\
\hline 
\normalsize{$\dot{E}_{GW}$ (in percentage)} & 
\normalsize{5.55} & 
\normalsize{upper limit}  	\\
\hline 
\normalsize{$\dot{E}_{em}$ (in percentage)} & 
\normalsize{94.44} & 
\normalsize{lower limit}  	\\
\hline
\end{tabular}
\caption{Upper and lower limits of energy ratio $Y_{n}$ between gravitational radiation and electromagnetic braking, electromagnetic braking index $n_{em}$ and upper and lower limits of gravitational and electromagnetic radiation contributions $\dot{E}_{GW}$ and $\dot{E}_{em}$ to the total energy loss.}
\label{Crab1}
\end{table}
\subsection{Energetics of the Crab pulsar's spin-down}
A large body of literature discusses the topic of the energy budget of the Crab pulsar and the surrounding Crab Nebula. See Ref. \cite{c12}, \cite{c16} for comprehensive reviews. The wind nebula acts as an effective calorimeter that allows us to measure the energy interactions between the pulsar and its environment. The mechanical spin-down is considered the main power source for both the pulsar and the Nebula luminosity. This power is about $4.77\times10^{38}$ ergs $s^{-1}$. The pulsar total luminosity in different electromagnetic wavelengths accounts just for a few percent of the total observed spin-down. For example, the pulsar radio luminosity is about $6.9\times10^{31}$ ergs \cite{c14} or $1.5\times10^{-7}$ of the mechanical loss of energy, while the X-ray luminosity is about $10^{-3}$. The total luminosity of the Nebula is about $1.3\times10^{38}$ ergs $s^{-1}$. It is well established that the main mechanism that allows the pulsar to transfer a large part of its mechanical energy to the Nebula is synchrotron radiation. For this process an efficiency of about 30 percent is sufficient to explain the observed total Nebula's luminosity. The remaining 70 percent of the energy loss is somehow not showing up in either the Nebula or the pulsar electromagnetic spectrum. In the mixed braking model that was discussed in the previous section, the upper limit on the efficiency of the gravitational braking was about 6 percent.  Therefore, this value of the efficiency is consistent with the Crab pulsar's and Nebula's energy budget as based on current observations. Instead, the assumption usually adopted to calculate the spin-down limit (all the energy loss is due only to gravitational waves) is, of course, in flagrant contrast with the above arguments. In fact, a large amount of energy is released into the Nebula. This energy cannot be in the form of gravitational waves that do not interact readily with matter.\\ The Crab pulsar's energy loss due to gravitational waves, implied by the most recent upper limit from LIGO, is now more consistent with other astrophysical observations, as the pulsar and Nebula electromagnetic spectrum. The implied upper limit on ellipticity itself is still too large from a theoretical point of view. The current value $\epsilon=1.8\times10^{-4}$ is at best at the limit of what the most exotic neutron star models allow \cite{c15}.\\As better limits on the gravitational strain of the Crab pulsar will be set by future LIGO data, we will have tighter limits on the efficiency of the gravitational wave contribution to the spin-down. At the same time, the contribution of the gravitational braking to the overall energy loss will become quickly less quantitatively significant as the efficiency depends on the square of the ellipticity. An improvement of a factor of 10 in strain sensitivity (easily reachable by Advanced LIGO) would mean an improvement on the upper limit on ellipticity of the same order. In this case, the upper  limit on the efficiency of the gravitational wave braking will be about 100 times smaller than the current one. At that point, the gravitational wave luminosity upper limit will be comparable with the pulsar's X-ray luminosity (and yet about ten thousand times higher than the radio luminosity). See Fig. 2.

\begin{figure}
	\centering	\includegraphics[width=0.45\textwidth]{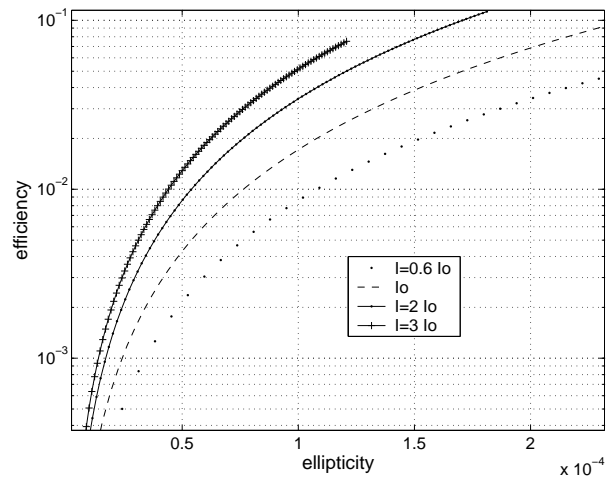}
	\caption{Efficiency of the Crab pulsar's gravitational radiation braking versus ellipticity. The ellipticity is in the range between the upper limit given by the current LIGO observations and possible limits given by future observations. The different curves represent different values of the moment of inertia. $I_{o}=1\times10^{45} g cm^{2}$ is the canonical value for the moment of inertia. The ellipticity and a neutron star's moment of inertia are dependent parameters given a certain observational upper limit on the gravitational strain. See section III.}
	\label{fig:MdotBeta1}
\end{figure}

\section{A more complete model with $n<3$}
The dipole radiation model is the most traditional braking mechanism suggested \cite{c4} and one of the most successful in explaining many of the pulsars' observed properties.\\
An interesting alternative model is proposed by Chan and Li (2006) \cite{c7}. The authors assume in their model a conventional dipole torque and, to make the overall braking index consistent with a value less than 3, a fallback disk torque is invoked. It is possible to add a gravitational torque produced by an asymmetry in the crust of the Crab pulsar as given by the LIGO ellipticity upper limit. In this case, the energy loss can be described by the following equation:
\begin{equation}
\dot{E}=I\Omega\dot{\Omega}=I\Omega\left(\dot{\Omega}_{GW}+\dot{\Omega}_{em}+\dot{\Omega}_{Td}\right), 
\end{equation}
where the energy loss contribution due to the fall back disk can be obtained from:
\begin{equation}
I\dot{\Omega}_{Td}=-\gamma^{-1}c^{-1/4}R^{9/4}\beta\left(3\alpha/4\right)^{1/2}\dot{M}\Omega^{7/4}t^{1/2},
\end{equation}
where $c$ is the speed of light, $R$ is the radius of the neutron star, $t$ is the historical age of the star, $\beta$ is the pressure scale height to radius ratio, $\gamma$ is the light-cylinder radius of the pulsar and $\alpha$ is the viscosity parameter. As in \cite{c7} we use, for our calculations in this letter, the following values for these parameters: $\beta=0.35,0.45,0.55$, $\gamma=1$ and $\alpha=0.001$. The parameter $\dot{M}$ is the mass inflow rate (assumed to be constant). Once the other parameters are set we are going to calculate lower limits on the parameter $\dot{M}$ for a given value of the free parameter $\beta$. Using equation (15) it is possible to show that the braking index $n$ can be rewritten as in the following:
\begin{equation}
n=\frac{\ddot{\Omega}\Omega}{\dot{\Omega^{2}}}=3-2\eta_{GW}-\left(\frac{5}{4}+\frac{\tau}{t}\right)\eta_{T_{d}}.
\end{equation}
where $\tau=-\Omega/(2\dot{\Omega})$ is the canonical age of the pulsar, $\eta_{GW}=\dot{\Omega}_{GW}/(I\dot{\Omega})$ is the efficiency of the gravitational torque and $\eta_{T_{d}}=\dot{\Omega}_{Td}/(I\dot{\Omega})$ is the efficiency of the disk braking torque.
Equation (15) and (17) can be solved as a system to find unique values for the lower limits on the magnetic field $B$ of the pulsar and the fallback disk mass inflow rate $\dot{M}$ given the parameter $\beta$ and the upper limit on ellipticity of the Crab pulsar derived from the LIGO null observation.\\To further our analysis it is important to point out that in actuality the ellipticity $\epsilon$ and the moment of inertia $I$ are not independent parameters in our model. In fact, the gravitational strain upper limit sets an equivalent upper limit only on the quadrupole moment $I\epsilon$. It is then customary to use a canonical value of $I=10^{45}$ g cm$^{2}$ to calculate an upper limit on ellipticity. But the possible values for $I$,  suggested by theory or observation, span a range within a factor of few from the canonical value \cite{c18}. Therefore, we express our results in terms of dependency on the moment of inertia $I$.
Figure 3 shows the mass inflow $\dot{M}$  as a function of the moment of inertia $I$ for different realistic values of the parameter $\beta$. For example, for a value of $\beta=0.035$ and a canonical value of $I$ we have $B=1.07\times10^{12}$ gauss and $\dot{M}=2.87\times10^{18}$ $g s^-1$.

\begin{figure}
	\centering	\includegraphics[width=0.45\textwidth]{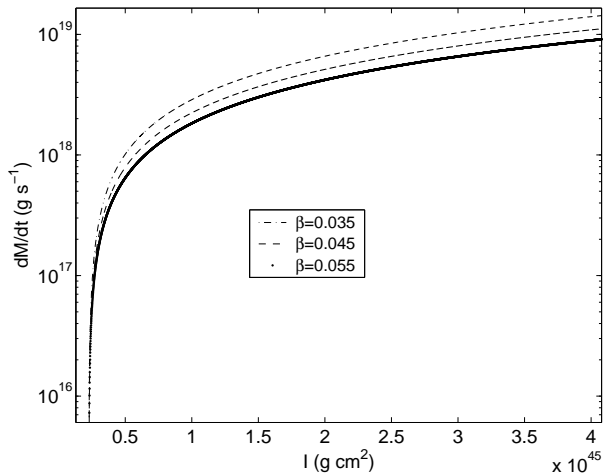}
	\caption{The fallback disk mass inflow rate $\dot{M}$ as a function of th moment of inertia $I$ of the pulsar and the dimensionless parameter $\beta$.}
	\label{fig:MdotBeta1}
\end{figure}

\begin{figure}
	\centering	\includegraphics[width=0.45\textwidth]{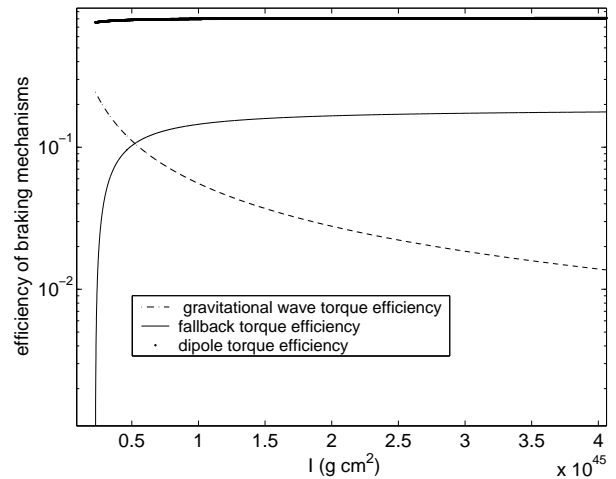}
	\caption{The efficiency of the different breaking mechanisms for the fallback disk model as a function of the pulsar moment of inertia $I$. Small moments of inertia correspond to large efficiency of the gravitational torque. The limiting value of $I=2.2\times10^{44}$ g cm$^{2}$ corresponds to the smallest value of $I$ consistent with the presence of a fallback torque. }
	\label{fig:efficiency}
\end{figure}

Figure 4 compares the efficiencies of the different braking mechanisms as a function of the angular momentum $I$. The greatest efficiencies for the gravitational wave braking is given by small values of $I$. Notice that there is a limiting value of about $I=2.2\times10^{44}$ g cm$^{2}$ that still allows for the presence of a gravitational wave torque and the existence of a fallback disk. Values of $I$ smaller than this give an unrealistic negative mass inflow.\\ These are just examples of how upper limits based on gravitational wave observations can start to constrain astrophysical models of neutron stars. For what concerns the energetics of this model, using the canonical value of the moment of inertia, we have that the electromagnetic braking efficiency is equal to about 80 percent, the gravitational braking efficiency is 6 percent and the efficiency of the disk braking torque is about 14 percent. Given the large ellipticity assumed in this particular model the gravitational torque is actually comparable in magnitude to the fallback disk torque.

\section{Conclusions}
The reaching and beating of the Crab pulsar's spin-down upper limit is an important milestone for the LIGO project. The current strain sensitivity upper limit can be converted to give an upper limit for the ellipticity of $1.8\times10^{-4}$.\\ This upper limit for ellipticity is still too large to be realistic from an astrophysical point of view. However, we can start to make interesting astrophysical statements on the braking mechanisms and energetics at play in slowing down the star.\\
In fact, we can now free the constraint that all the energy loss is due to the emission of gravitational radiation and assume a mixed braking mechanism with contributions from different physical phenomena (gravitational radiation, electromagnetic dipole radiation, winds, magnetospheric currents, fallback disk torques, etc). \\Different models can be adopted. One of the models we have studied assumes that the electromagnetic braking index is different from the conventional $n=3$. We can use the upper limit on the ellipticity to calculate the contribution to the energy loss due to gravitational radiation and calculate the parameter $n_{em}$, the braking index due to electromagnetic processes. Our current result for the Crab pulsar is that the ratio between the energy loss due to gravitational radiation to the energy loss due to electromagnetism is 0.058. This implies that the contribution of the gravitational radiation to the total energy loss has to be less than 6 percent. Finally, the electromagnetic braking index is 2.36 (this is a lower limit). \\Alternatively, we can use a model that allows for a classical electromagnetic dipole but assumes other braking mechanisms to account for a overall breaking index different from $n=3$.
In this case we can show that the efficiency of the gravitational wave braking (given the current limits on ellipticity from LIGO) is comparable to that of a fallback disk torque with realistic parameters. The limit on ellipticity can be used to set limits on the mass inflow for the fall back disk. It would be necessary to have a mass inflow $\dot{M}$ of about $ 2.87\times10^{18}$ $g s^-1$ to make the model consistent with the limit on ellipticity.
Even more interesting conclusions on the limits of the braking mechanisms and energetics of pulsars will be reached with advanced detectors.


\bibliographystyle{unsrt}
\bibliography{CrabSpinDownLimit2}
\end{document}